# THE 2011 OCTOBER DRACONIDS OUTBURST. I. ORBITAL ELEMENTS, METEOROID FLUXES AND 21P/GIACOBINI-ZINNER DELIVERED MASS TO EARTH

**Josep M. Trigo-Rodriguez[1,2], José M. Madiedo[3,4], I. P. Williams[5], Joan Dergham[1], Jordi Cortés[1], Alberto J. Castro-Tirado[6], José L. Ortiz[6], Jaime Zamorano[7], Francisco Ocaña[7], Jaime Izquierdo[7], Alejandro Sánchez de Miguel[7], Jacinto Alonso-Azcárate[8], Diego Rodríguez[9], Mar Tapia[10], Pep Pujols[11], Juan Lacruz[12], Francesc Pruneda[13], Armand Oliva[14], Juan Pastor Erades[14], and Antonio Francisco Marín[15].**

[1] Institute of Space Sciences (CSIC), Campus UAB, Fac. Ciències, Torre C5-par.2ª, 08193 Bellaterra, Barcelona Spain.
[2] Institut d'Estudis Espacials de Catalunya (IEEC), Edif.. Nexus, c/Gran Capità, 2-4, 08034 Barcelona, Spain
[3] Facultad de Ciencias, Universidad de Huelva, Huelva, Spain.
[4] Departamento de Física Atómica, Molecular y Nuclear. Facultad de Física. Universidad de Sevilla. 41012 Sevilla, Spain.
[5] School of Physics and Astronomy, Astronomy Unit, Queen Mary, University of London, London, UK.
[6] Instituto de Astrofísica de Andalucía (IAA-CSIC), Granada, Spain,
[7] Depto. de Astrofísica y CC. de la Atmósfera, Facultad CC Físicas, Universidad Complutense de Madrid (UCM), Madrid, Spain,
[8] Universidad de Castilla-La Mancha (UCLM), Toledo, Spain
[9] Guadarrama Observatory (MPC458), Madrid, Spain.
[10] Laboratori d'Estudis Geofísics Eduard Fontseré (LEGEF), Institut d'Estudis Catalans, Barcelona, Spain,
[11] Grup d'Estudis Astronòmics (GEA) and Agrupació Astronòmica d'Osona, Barcelona, Spain.
[12] La Cañada Observatory, Ávila, Spain,
[13] Astronomia des de l'Empordà, Palamós, Spain,
[14] Agrupació Astronómica de Sabadell (AAS), C/Prat de la Riba s/n; 08200 Sabadell, Barcelona, Spain.
[15] Urb. Villapalma 10, 1º D, 11203 Algeciras, Cádiz, Spain.

**ABSTRACT**: On October 8th, 2011 the Earth crossed the dust trails left by comet 21P/Giacobini-Zinner during its XIX and XX century perihelion approaches with the comet being close to perihelion. The geometric circumstances of that encounter were thus favorable to produce a meteor storm, but the trails were much older than in the 1933 and 1946 historical encounters. As a consequence the 2011 October Draconid display exhibited several activity peaks with Zenithal Hourly Rates of about 400 meteors per hour. In fact, if the display had been not forecasted, it could have passed almost unnoticed as was strongly attenuated for visual observers due to the Moon. This suggests that most meteor storms of a similar nature could have passed historically unnoticed under unfavorable weather and Moon observing conditions. The possibility of obtaining information on the physical properties of cometary meteoroids penetrating the atmosphere under low-geocentric velocity encounter circumstances motivated us to set up a special observing campaign. Added to the Spanish Fireball Network wide-field all-sky and CCD video monitoring, other high-sensitivity 1/2" black and white CCD video cameras were attached to modified medium-field lenses for obtaining high resolution orbital information. The trajectory, radiant, and orbital data of 16 October Draconid meteors observed at multiple stations are presented. The results show that the meteors appeared from a geocentric radiant located at $\alpha=263.0\pm0.4°$ and $\delta=+55.3\pm0.3°$ that is in close agreement with the radiant predicted for the 1873-1894 and the 1900 dust trails. The estimated mass of material from 21P/Giacobini-Zinner delivered to Earth during the six-hours outburst was around 950±150 kg.

1. Introduction: The 2011 October Draconid outburst.

Meteor storms are rare, but magnificent displays of nature that remind us of the crucial role that the terrestrial atmosphere can play in shielding us from direct impacts by interplanetary particles. Meteoroids with sizes over ~100 µm typically ablate in the atmosphere where some of the kinetic energy generates a visible trail, a meteor (McKinley, 1961). Because of the effects of perspective, when observed from the ground meteors seem to fall in their hundreds over a very short timescales of minutes or even seconds (Fig. 1). As well as being spectacular, the study of meteor storms can be of great scientific value. From multi-station recordings of meteors the velocity, deceleration, and dynamic strength of the meteoroid can be measured, while from the radiant and the deduced velocity their heliocentric orbits can be calculated.

From an astrobiological perspective, the encounter of our planet with dense meteoroid streams under favorable geometric circumstances can also provide a unique opportunity to quantify the delivery of volatile-rich materials to Earth. At the present time, these rare encounters represent a sample of the delivery of organic molecules and water that were common in the past. These mechanisms could have participated in the terrestrial enrichment in volatiles at the time of the Late Heavy Bombardment (Gomes et al., 2005). At that epoch fragile bodies were scattered by Jupiter and Saturn from the Kuiper Belt and the outer Main Belt disk, crossing the orbits of the terrestrial planets and experiencing regular close encounters. Direct impacts probably occurred, and could have been a key source of volatiles to Earth, but fragmentation of these ice-rich bodies in dust trails could have open additional pathways (Trigo-Rodríguez & Martín Torres, 2013).

Though ancient written records of meteor storms are common, it is difficult to infer the fluxes of meteoroids to Earth from such historical reports because until the $20^{th}$ century meteor observing was not standardized (Jenniskens, 2006). The best ancient meteor reports provided the hourly rates but with hardly any information on cloud cover or the state of the moon. Despite the difficulties of interpreting the observations, past civilizations observed the skies in far better sky conditions than we do, but they were not able to understand what was being observed. Now, the meteor rate registered hourly by visual observers is standardized as the Zenithal Hourly Rate (ZHR). The ZHR corrects for several effects such as the zenith distance of the radiant, the stellar limiting magnitude (Lm) and the percentage of sky covered by clouds.

With modern techniques, the recording of meteor storms and computing the ZHR can be carried out to a level of high accuracy. In general, a shower is called a meteor storm if the ZHR exceeds 1,000 meteors per hour. In contrast, the sporadic meteor rate is usually less than 10 per hour. In perfect conditions, with the radiant at the zenith, no obstacles and +6.5 limiting stellar magnitude with the naked eye, such a ZHR corresponds to a meteor frequency of about 1 meteor every 4 seconds. Such a rate will produce an obvious meteor display that can be seen even by inexperienced sky observers. Several comets are known to produce meteor storms, and 21P/Giacobini-Zinner is one of them. On Oct. $9^{th}$, 1933, a Catalan astronomer Josep Comas Solà observed one of the most intense storms and described it in a famous popular book: "from the beginning of the night until 22h, at least tens of thousands of meteors were observed over all Europe" (Comas Solà, 1939).

Until the arrival of modern computers, meteor storm forecasting was a difficult task. Meteor storms are produced by tiny particles with typical sizes of tens or hundreds of microns that were released from a comet nucleus. The emission is driven by the sublimation of ices (Whipple, 1951). As the orbits of the dust particles differ by a small amount from that of the parent comet, they will have slightly different orbital periods so that over time meteoroids will spread all around the orbit so that a meteor shower can be observed every year. (For a description of all the physics and mathematics involved in the process, see for example Williams, 2002, or a shorter version in Williams, 2004). However such spreading takes time and meteoroids ejected from the parent in the last few hundred years will still be preferentially clumped close to the comet location on its orbit. Meteor storms can thus be expected when the stream is young at the time where the comet is close to perihelion (see Williams 1997 for a discussion of this process). The appearance of a storm however depends both on how close the nodal distance of the stream orbit is to the Sun-Earth distance and on how short the time interval is between the meteoroid clump passing through the node and the Earth reaching the same point as was shown by Wu & Wiliams (1995, 1996). This principle was also used by Asher (1999) to explain the Leonid storms.

Comet 21P/Giacobini-Zinner is one comet that is capable of producing a dense meteoroid stream which can result in a fantastic meteor display when the Earth passes through the centre of the stream close. Such storms were seen in 1933 and 1946 as the Draconid meteor storms (also historically called the Giacobinids) when rates went up to ZHR=10,000 (Jenniskens, 2006). Such meteor displays are among the strongest storms ever seen. The circumstances both in terms of nodal distance and time between the comet being at its node and the Earth passing this point were predicted to be similar again in 2011 (Jenniskens, 2006; Maslov, 2011; Vaubaillon et al., 2011). All the models predicted that on October 8.7, 2011 the Earth would encounter the dust trails left by comet 21P/Giacobini-Zinner during its 19th and 20th century perihelion approaches.

It is important to remember that the spatial density of meteoroids in streams can decrease with time. Both planetary perturbations due to close encounters with planets (Hughes Williams & Fox, 1981; Jenniskens, 1998) and mutual collisions among particles from the same or different dust trails can contribute to remove meteoroids from the stream (Babadzhanov et al, 1991; Williams et al, 1993; Jenniskens, 1998). Collisions with Zodiacal dust particles can also cause vaporization or fragmentation (Trigo-Rodríguez et al., 2005). Hence, even if the encounter circumstances are identical, the above effect can reduce the observed strength of a storm.

The 21P dust trails were obviously older in 2011 than in either 1933 or 1946 and so were expected to be less dense. The decreased flux number density expected for the comet trails was confirmed as during the 2011 Draconid shower the ZHR was less than 1000, making it technically an outburst rather than a storm. The storm was also not so visible to the general public, because of the previously noted presence of the Moon (Jenniskens, 2006; Babadzhanov et al., 2008). However, modern recording systems can work well even in non-favorable conditions, and so a special observing campaign was initiated for the 2011 Draconids. For this, a multistation CCD and video monitoring systems of the Spanish Meteor Network (SPMN) were used. An additional amateur campaign was also initiated. The SPMN high sensitivity CCD allowed reliable flux and orbital information on the meteoroids that produced the outburst to be obtained. In addition, a −10.5±0.5 absolute magnitude Draconid bolide over Andalusia, Spain was observed implying that the original meteoroid had a mass of about 13 kg (Madiedo et al., 2013). Large fireballs also produced long-lasting persistent trains, and some examples of such spectacular phenomena are given.

This paper has three main goals. First, to summarize the results on the meteoroid flux at the Earth from 21P dust trails derived from visual, video and radio stations during October 2011. Second, to present the trajectory, radiant, and orbital data of the most precise orbits computed so far by the SPMN. Third, to compare observational data with the theoretical forecasting in order to provide information on the small-scale structure of the 21P dust trails. This information will be of use for future forecasting of Earth's encounters with cometary dust trails, particularly to better quantify the effects of aging processes in meteoroid streams.

2. Instrumentation, Data Reduction and Observation sites.

Trigo-Rodríguez et al., (2004) have already outlined the first steps in the development of the SPMN that use low-scan-rate all-sky CCD cameras with +2/+3 meteor limiting magnitude. In 2006 a further expansion of the network took place when two new all-sky CCD stations in Catalonia and three video stations in Andalusia were added. There are now 25 stations distributed all over Spain. The main goal of the monitoring project is to increase the observations of meteor and fireball activity from multiple stations (Trigo-Rodríguez et al., 2006, 2007). The SPMN stations use high-sensitivity CCD, and video cameras to monitor the night sky. The video cameras are equipped with a 1/2" Sony interline transfer CCD image sensor with their minimum lux rating ranging from 0.01 to 0.0001 lx at f1.4 (Madiedo and Trigo-Rodríguez, 2007). Aspherical fast lenses with focal length ranging from 4mm (fisheye) to 25 mm and focal ratio between 1.2 and 0.8 are used for the imaging objective lens that typically reach a limiting magnitude of +4. In this way, different areas of the sky can be covered by every camera and point-like star images are obtained across the entire field of view. The observing stations are automatically switched on and off at sunset and sunrise, respectively. The cameras generate video imagery at 25 frames per second with a resolution of 720x576 pixels$^2$ and are continuously sent to PC computers through a video capture card. Computers execute software (UFOCapture, by SonotaCo) for automatic detection of meteors and storage of the corresponding frames on hard disk. Since the time of a meteor appearance is crucial in orbital determination, the computers are synchronized by means of GPS devices. In this way, the time is measured with an accuracy of $10^{-1}$ s along the entire meteor path.

Astrometric reduction of imagery is performed using software described elsewhere (Trigo-Rodríguez et al., 2003; Madiedo Trigo-Rodríguez & Lyytinen, 2011). In any meteor event, the software obtains a composite image where automatic detection of stars is achieved. The stars are then measured one at a time and those with significant S/N ratios selected for astrometric reduction. Note that no software is used for automatic astrometry of the images so that the observer performs the precise astrometry for stars in the composite image and for the meteor moving in each individual frame. It is then necessary to identify meteors that are common to several observing stations. Under normal meteor activity circumstances, a preliminary search through the database of meteors that appeared during the same observing interval produces the unequivocal identification of common multiple-station meteors if GPS time calibration is performed in all stations. An interesting application in our software packages is particularly useful for meteor storms, namely the ability to predict the position of every meteor from each station once the astrometry from one station is completed and assuming the typical values of ablation height. The astrometric measurements from each station are then introduced into our *Network* and *Amalthea* software packages (Trigo-Rodríguez et al., 2003, Madiedo et al., 2011), which compute the equatorial coordinates of the meteors with an astrometric accuracy of about 0.01º and also determine the apparent and geocentric radiant of common meteors. Once identified, from the measured sequences recorded in two or more stations, the software estimates by triangulation the atmospheric trajectory and radiant for each meteor.

It is important to explain and quantify the errors in the results. The accuracy of the astrometric data is directly measured from the standard deviation of the background stars compared with the meteor positions as explained in Trigo-Rodríguez et al. (2003). From the inferred beginning and ending meteor co-ordinates from both stations, and their respective standard deviation uncertainties the radiant location is obtained. Then the astrometric accuracy propagates into a standard deviation in the radiant position for each meteor as given in Table 5. We selected favorable cases for astrometric reduction, except when the fields of view are wide and slightly distorted due to spherical aberration. Even when correction of that effect has been implemented following the approach by Steyaert (1990) instrumental scattering is still noticeable in the radiant data as shown in Fig. 3. We suspect that this effect could be due to the pixel size of the detector in which the meteor image is focused and becomes larger as the distance between the meteor and the apparent radiant increases, so the best way to deal with it is probably measuring a large number of meteors to attenuate statistically the scattering. In fact, the averaged geocentric radiant fits well the expected theoretical position as is explained in the discussion.

Finally, in order to determine orbital elements from our trajectory data we used the *Amalthea* program that provides reliable trajectory, physical properties and orbital data.

3. Observational results: Spatial fluxes, trajectory, radiant and orbital data.

3.1. Determination of population index and meteoroid spatial fluxes.

Because they provide photon counts for every pixel, CCD cameras allow a very accurate determination of stellar and meteor magnitudes to be made. In all-sky CCD imaging a simplistic approach is adopted whereby meteor magnitudes are derived by comparing the intensity level of the pixels near the maximum luminosity of the meteor trail with those of nearby stars. The different angular velocity of the meteors should be taken into account as a function of the distance to the radiant and the typical duration of flares, but in general for meteors a difference of four magnitudes is produced, i.e. a meteor of magnitude -2 exhibits a path with similar intensity to a star of magnitude +2. General formulae to take into account the different angular velocity of the sources (stars and meteors) were compiled by Rendtel (1993). This generalization is not valid for meteors that appear below 30º of altitude since they need to be additionally corrected for atmospheric extinction losses, that we also corrected. Our measured magnitudes were additionally tested for correctness to within ±0.5 magnitudes by performing simultaneous visual observations and correlating the meteor peak to the imaging record.

From the visual and video derived meteor magnitudes, the magnitude distribution for the night of Oct. 8-9, 2011 was obtained and is given in Table 2. From this a population index for the three experienced visual observers of r=2.3±0.3 (N=393) was derived. This value was used to estimate the visual Zenithal Hourly Rate (ZHR) as well as to convert to the spatial flux of meteoroids meteors brighter than +6.5 per km$^2$ given in Table 3. The results suggest at least two peaks with a maximum visual (human) rate close to ZHR=400. In general, the values confirm the visual rates compiled by amateurs in the framework of the International Meteor Organization (IMO webpage). A general discussion of the results presented in Tables 3 and 4 is particularly useful to understand the interception of 21P dust trails by Earth. Observations can be compared with the excellent forecast of the interception of the 21P dust trail by Earth made in Table 3 of Vaubaillon et al. (2011). The determined flux in the ]-∞, +5] magnitude range was maximum at solar longitude 195.0106º (Oct. 8$^{th}$, 2011 at ~19h38m UTC) when the flux reached $(113\pm16)\times10^{-3}$ km$^{-2}$ h$^{-1}$. This peak fits perfectly, particularly taking into account the arbitrary periods taken, with the time forecast for the 1907 dust trail at solar longitude 195.0059º (Vaubaillon et al., 2011). A second peak occurs at solar longitude 195.0311º (Oct. 8$^{th}$, 2011 at ~20h08m UTC) when the visual flux reached $(102\pm13)\times10^{-3}$ km$^{-2}$ h$^{-1}$. This second peak also agrees with that forecasted by Vaubaillon et al. (2011) for the dust trail released by comet 21P during the 1900 perihelion passage. Finally, a third peak of similar intensity occurred at solar longitude 195.0721º (Oct. 8$^{th}$, 2011 at ~21h08m UTC) when the visual flux reached $(106\pm16)\times10^{-3}$ km$^{-2}$ h$^{-1}$. That peak also produced bright meteors, and may be the result of several older dust trail components as it is not clearly predicted in Vaubaillon et al. (2011). The visual comparison among visual and video data shows that the third peak was not recorded in video observations (Fig. 2a). Despite this, a moderate peak at that solar longitude is seen in IMO data (IMO webpage), but the absence in our video records perhaps supports the idea that this peak was mainly composed of faint meteors as suggested by the decreasing population index values (see Fig. 2b). The existence of this was confirmed by backscatter radio observations (see radio counts in Table 4) with three consecutive 10-min. intervals exhibiting high rates around 21h00m UTC. A discone antenna was used together with a Yaesu VR5000 receiver working at 143.05 MHz from Guadarrama Observatory (Madrid). This radio data seems to reveal more moderate radio bursts at 17h15m, and 17h45m UTC probably associated with the 1887 dust trail, and another one at 19h05m that could be tentatively associated with the 1894 dust trail (Table 3 of Vaubaillon et al., 2011). The Giacobinid flux was about one order of magnitude lower for bright meteors recorded by video cameras with +3 limiting magnitude that night. Consistently, the corrected SPMN counts were found to be 40 times stronger during the outburst than for sporadic rates that usually reach ~10 meteors/hour.

To roughly compute the amount of mass delivered by comet 21P to Earth during the 2011 outburst ($M_{DEL}$) we use a first order of magnitude approach. ($M_{DEL}$) is computed by considering the number of meteoroids in each magnitude range and multiplying them by the meteoroid mass given in Appendix C, equation C.12 of Jenniskens (2006). The number of meteoroids in each magnitude range is fitted to be what is required to produce an averaged global ZHR of ~400 with a population index: r~2 (Table 3). As the ZHR was slightly lower than that in most intervals, our computation is an upper limit for the mass delivered. According to the radio data shown in Table 4, the outburst level was sustained for about 6 hours, and that value was used for the final computation. The equations that describe the procedure are:

$$M_{DEL} = \sum_{i}^{+6} m_i \cdot N_i \quad where \quad \sum_{i}^{+6} N_i = ZHR(r)_{observed}$$

This gives the meteoroid mass delivered into a subtended atmospheric volume seen by a visual observer (see Koschack & Rendtel J., 1990). This mass needs to be multiplied by a factor to cover the total mass reaching all Earth. The final result of such calculations yields $M_{DEL}$=950±150 kg delivered during the six hours of outburst. To obtain the mass uncertainty we applied the common law of propagation of errors having into account the uncertainty in the meteor flux and the population index. Obviously, a significant part of this mass was ablated during atmospheric interaction, but contributed significantly to the release of elements in the upper atmosphere.

### 3.2. Trajectory, dynamic strength and radiant data.

The observed common field for the stations was initially programmed (Section 2) so that double-station meteors were required to have convergence angles greater than 20º to allow accurate determination of trajectory and geocentric radiant. The convergence angle (Q) is the angle between the two planes delimited by the observing sites and the meteor path in the triangulation. The trajectory data of 16 accurately reduced meteors are given in Table 5, which shows the SPMN code used for identification, the apparent visual magnitude ($M_v$), the meteor trail beginning and end height on the Earth's surface ($H_b$ and $H_e$ in km), the geocentric radiant coordinates ($\alpha_g$ and $\delta_g$ to Eq. 2000.00) and the velocity in km·s$^{-1}$ (at the top of atmosphere, geocentric and heliocentric). The velocity at the top of the atmosphere was measured in the upper parts of the luminous trajectories, and double-checking that the measured values adjust to the values derived for the following frames.

From the 16 Draconid radiants we obtained an averaged geocentric radiant at $\alpha$=263.0±0.4º and $\delta$=+55.3±0.3º. For comparison, the theoretical radiants given by Maslov (2011) or Jenniskens and Vaubaillon (2011) are compiled in Table 6. The orbital parameters are given in Table 7. The radiant and average velocity data based on the data in Table 5 are in close agreement, but far more precise, than those discussed in the IMO list by Langbroek (2011) from a joint American/German/Dutch video campaign to study the outburst (last row in Table 6). Finally in Figure 3 are shown the October Draconid geocentric radiants compared with the theoretical position given by Maslov (2011).

### 3.3. Orbital elements of 2011 October Draconid meteors.

From the radiant position, appearance time and velocities estimated for the Draconid meteors listed in Table 5 we derived the orbital elements shown in Table 7. Due to the high meteor rate, we decided to name the meteors from the appearance time (SPMN: hour:minute:second). This way is also useful to identify the probable dust trails to which the meteors or fireballs belong. For example, the first 8 meteors in Table 7 appeared in 1 hour interval from 8.78 to 8.81 Oct. 2011. Consequently, looking at Fig. 2 a, they are very likely associated with the 1907 dust trail. The 1900 dust trail detections start with the extraordinary −10.5 magnitude bolide SPMN194759 shown in Fig. 4 (see detailed study about its emission spectrum by Madiedo et al. (2013)). At that interval from 8.82 to 8.87 Oct. we computed high-resolution orbits of 8 bright meteors, six of them practically in the fireball range. This would suggest that a small fragmentation event could have taken place on 21P during its 1907 perihelion passage since it is not possible for meteoroids larger than about 10 cm to be ejected by the normal Whipple mechanism (Williams, 2004).

4. Discussion.

An important consequence of being able to obtain accurate trajectory data is that the physical properties of the meteoroid can be determined. Meteors that exhibited a catastrophic disintegration at the end of their paths allow their dynamic strengths to be determined (Trigo-Rodríguez & Llorca, 2006, 2007). This was the case for many of the Draconid meteors. To do this, the aerodynamic strength (S) is required and we have used the equation given by Bronshten (1981):

$$S = \rho_{atm} \cdot v^2 \qquad (1)$$

where $\rho_{atm}$ is the atmospheric density at the height where the meteoroid breaks up and $v$ is the velocity of the particle at this point to estimate this. If the density is given in kg/m$^3$ and the velocity in m/s, the strength is given in dyne/cm$^2$. Verniani (1969) and Wetherill & ReVelle (1982) applied this equation for determining mechanical stresses. Verniani (1969) pointed out those meteoroids following typical cometary orbits fragment when the pressure exceeds $2 \times 10^4$ dyn cm$^{-2}$.

In Table 8 we show the heights, velocities, and dynamics strengths for four Draconids. Three of them exhibited catastrophic disruptions so we computed the strength for those points, but in the case of the bright bolide SPMN194759 the strength was computed in the first major flare (as discussed in Madiedo et al., 2012). We have computed the dynamic strength for these three cases following the approach described in (Trigo-Rodríguez & Llorca, 2006, 2007). The Draconids appear to be the most fragile meteoroids from all the cometary showers with typical dynamic strengths below $\sim 10^3$ dyn/cm$^2$.

These results are consistent with the low strength cometary populations identified by Jacchia (1958) and Ceplecha (1958). On the other hand, Verniani (1969, 1973) and Millman (1972) found that most of the sporadic meteoroids of cometary origin are highly porous. In fact, cometary disruption events are occurring even at large heliocentric distances, characteristic of extremely crumbly structures (Sekanina, 1982). Such events provide clues to the extremely low tensile strengths of cometary nuclei, estimated to be between several times $10^3$ and $10^5$ dyn/cm$^2$ (Donn, 1963; Donn & Rahe, 1982). These measured strengths are consistent with the behavior of cometary meteoroids that typically fragment in the upper atmosphere at similar aerodynamic pressures. Further clues about the nature of cometary meteoroids can be obtained from the study of ballistic aggregation experiments (Krause & Blum, 2004). All this data suggests that cometary meteoroids may be fractal aggregates with extremely high porosity.

The study of the atmospheric interaction of cometary meteoroids penetrating the atmosphere at low geocentric velocities is also interesting from a cosmochemical point of view. From the changes in the population index and in the number of fireballs since these trails were crossed by Earth in the 1930s, we have evidence that Draconid meteoroids are being progressively eroded. The occurrence of such a progressive process occurs in the interplanetary medium may be explained in the context of the recent discovery of ultracarbonaceous micrometeorites in Antarctica (Duprat et al., 2010). Such fragile materials belong to some primitive parent bodies of isotopic and chemically exotic nature. For example, they exhibit high D/H ratios, abundant organic matter, and µm-sized or smaller silicate particles similar to these found in porous Interplanetary Dust Particles (IDPs). The delivery of biogenic elements by encounters with dense cometary trails along the eons probably has been relevant. Blum et al. (2006) reasoned from accretionary, dynamic and evolutionary arguments that hundred- to km-sized primitive asteroids and comets should exhibit a fragile nature: extremely low bulk density, and high porosity. Recent Stardust collection of cometary dust in the coma of pristine comet 81P/Wild 2 also provided interesting clues on the nature of these materials (Brownlee et al., 2006). They are aggregates whose structure is similar to carbonaceous IDPs or primitive carbonaceous chondrites. Consequently, we expect a 21P/Giacobini-Zinner cometary meteoroid structure composed of a matrix rich in C and other biogenic elements, and additional chondrules, and tiny and rarer refractory inclusions. Due to the relative low bulk density and large porosity of those aggregates, the tensile strength of 21P/Giacobini-Zinner meteoroids is much lower than for any known terrestrial mud or sandstone. This fragile nature explains the brilliant catastrophic disruptions that we typically observe in the upper atmosphere for cometary-origin bolides (Trigo-Rodríguez et al., 2007, 2009; Trigo-Rodríguez & Blum, 2009). In fact, about 20% of the large fireballs recorded by the Prairie Network ended up in a sudden overwhelming fragmentation that translates into a flare, and about 60% of the cases experienced one or several fragmentations along their path (Ceplecha et al., 1998). This occurs when the meteoroid feels an increasing dynamic pressure ($p=\rho \cdot v^2$) as it penetrates the atmosphere. When the loading pressure surpasses the material strength required for fragmentation the body breaks apart and, as consequence of the flight and shock wave shaking, disruption is imminent. Once disrupted, most of the fine grained material exposed to the frontal bowl shock is very efficiently vaporized, as meteor spectroscopy reveals that the material quickly reaches temperatures well over the sublimation point of silicates (Borovicka et al., 1993, 1994; Trigo-Rodríguez et al., 2003; Madiedo et al., 2013). On the basis of fireball spectroscopy, it is suspected that catastrophic disruptions can disperse dust back and far from the shock wave frontal region where bolide experiences higher temperatures (Trigo-Rodríguez and Martín-Torres, 2013). Ablation temperature is lower for low-entry velocity meteoroids, and particularly in these cases the exposure of the released materials to heat may not be identical. A catastrophic break up could move dust laterally, and generate turbulence. If so, there is room for a small percentage of the body to survive, as supported by the discovery of unmelted dust and small micrometeorite fragments that are slowly setting down towards the surface (Taylor et al., 2000; Genge, 2008; Duprat et al., 2010). Indirect evidence on the survival of small quantities of dust in meteor spectroscopy could be the presence of a continuum of radiation in meteor spectra, or the persistent trains observed for seconds or even minutes after the extinction of the fireball phase. In any case, meteoroid smokes produced by recondensation of vaporized minerals can also contribute, and the association needs to wait until achieving spectroscopy of much higher resolution and fast video cameras. In any case, proof that dust can survive was provided in high resolution spectra obtained during the reentry and ablation of the impact plumes produced on the Jovian atmosphere as a consequence of the impact of comet Shoemaker-Levy 9 in July 1994 at a velocity of 60 km/s (Fitzsimmons et al., 1996). These authors found that most of the light emission came from silicate grains ablated in the different phases, even in the case of a bolide produced by tens of meters-sized cometary fragments. On the other hand, fireball entry models not only can predict survival of silicate dust, but also of more friable compounds like e.g. organics in the internal structure. In this sense, Blank et al. (2001) have shown that asteroids and comets impacting the atmosphere of Earth are delivering small amounts of complex organics if the impact geometry and velocity are favorable to produce a moderate deceleration and setting of the materials in the atmosphere.

Another important aspect to consider is the thermal processing that affects the materials subjected to ablation in the fireball column. As a consequence of the heat associated with the collisions with atmospheric gases, meteoric minerals are ablated, vaporized and dissociated. Elemental lines and molecular bands are remarkable features in bolide spectra (Fig. 1). It is been demonstrated that most of the fireball chemistry behind radiating light can fit perfectly a thermodynamic equilibrium model (Borovicka, 2001; Trigo-Rodríguez et al., 2003). This behavior is probably consequence of the quick mixing of air and meteoric plasma promoted by the supersonic movement, meteoroid spinning, and subsequent induced turbulence around the bolide. It is important to remark, however, that the production of different gases can be avoided in environments with different chemistry and radiative flux.

5. Conclusions.

Despite the expectation created with the return of 21P/Giacobini-Zinner's dust trails to Earth's vicinity on 2011 October 8, the display was moderate compared to previous encounters. Dust trails left by the comet were precisely forecasted through the perihelion approach, and that achievement was in practice an excellent advantage to set up a special SPMN campaign with smaller fields of view than these used in usual fireball network patrol. The video CCD camera systems whose excellent performances for meteor recording were initially described in Madiedo and Trigo-Rodríguez (2004) are again showing its potential with the current data. In spite of the moderate Draconid activity, our camera systems were able to record hundreds of meteors all over Iberian Peninsula by using high-sensitivity 1/2" black and white CCD video cameras (Watec, Japan) and 1/3" progressive-scan sensors attached to modified short-field lenses. We have presented the main results on the orbital and flux data obtained by SPMN camera systems on that night. Unfortunately the meteor shower did not reach storm category; but the outburst was really remarkable with peaks of activity of several hundreds of meteors per hour. As the meteor activity was predicted in advance special camera systems were set up that were able to cover a wide area network, permitting the collection of very valuable information for optical meteors as faint as magnitude +3. The findings obtained from the 2011 Giacobinid campaign are:

a) On 2011 October 8 the Earth encountered the dust trails left by comet 21P/Giacobini-Zinner during its XIX and XX century perihelion approaches. The trails were older than in previous 1933 and 1946 historical encounters, and significantly perturbed by Earth's encounters so they produced an outburst, but not a storm.

b) Video observations allow the physical behavior of cometary meteoroids penetrating the atmosphere at low geocentric velocity to be studied. Terminal catastrophic flares are typically produced at dynamic strength pressures over 400 Pa, but for largest meteoroids that can reach about 1 kPa.

c) SPMN averaged geocentric radiant data in R.A.=263.0±0.4º and Dec.=+55.3±0.3º fits the theoretical radiant well inside the astrometric uncertainties.

d) The above mentioned previous encounters also decreased the meteoroid spatial flux. This is probably a direct consequence of gravitational scattering of the dust trail individual members during such encounters, but it is also probable consequence of a fragile nature of meteors that, having low strength and fractal-like structure, are more exposed to direct collisional erosion (with Zodiacal dust or same-stream meteoroids) and also to solar irradiation. Both space weathering processes are probably decreasing the spatial number density of meteoroids in timescales of few centuries.

e) The 2011 Giacobinid flux rates were about one order of magnitude lower than expected. The global mass of 21P cometary materials delivered to Earth was $M_{DEL}$=950±150 kg. To improve future models, precise flux determinations as these presented here could be the key to better understand interplanetary space erosive processes, and their direct effect in the diffusion of dust trails, and meteor displays.

**Acknowledgements**

We are particularly grateful to all amateur observers that contributed to this study. We acknowledge support from the Spanish Ministry of Science and Innovation (projects AYA2009-13227, AYA2009-14000-C03-01 and AYA2011-26522), Junta de Andalucía (project P09-FQM-4555) and CSIC (grant #201050I043). We also thank the Draconid Recerca en Acció project (granted by Generalitat de Catalunya) in order to promote a cooperative amateur campaign in Catalonia. We also thank Dr Margaret Campell-Brown for many useful suggestions for improving this paper.

# TABLES

Table 1. SPMN stations involved in the Giacobinid high-resolution campaign. Acronyms for the different imaging systems are: AS (low-scan-rate CCD all-sky camera), WF (low-scan-rate CCD wide-field camera), and WFV (Wide field video cameras).

| Station # | Station (Province) | Longitude | Latitude (N) | Alt. (m) | Imaging system |
|---|---|---|---|---|---|
| 1 | Montsec, OAdM (Lleida) | 00º 43´ 46" E | 42º 03´ 05" | 1570 | AS |
| 2 | Montseny (Girona) | 02º 31´ 14" E | 41º 43´ 17" | 300 | WFV |
| 3 | Folgueroles (Barcelona) | 02º 19´ 33" E | 41º 56´ 31" | 580 | WFV |
| 4 | Seville (Seville) | 05º 58´ 50" W | 37º 20´ 46" | 28 | WFV |
| 5 | Cerro Negro (Seville) | 06º 19´ 35" W | 37º 40´ 19" | 470 | WFV |
| 6 | El Arenosillo (Huelva) | 07º 00´ 00" W | 36º 55´ 00" | 30 | AS+WFV |
| 7 | El Picacho (Cádiz) | 05º 39´ 01" W | 36º 31´ 19" | 392 | WFC |
| 8 | Madrid-UCM (Madrid) | 03º 43´ 34" W | 40º 27´ 03" | 640 | WFC |
| 9 | Villaverde del Ducado (Guadalajara) | 02º 29´ 29" W | 41º 00´ 04" | 1,100 | WFC |
| 10 | Toledo | 03º 57´ 29" W | 39º 49´ 30" | 639 | WFC |
| 11 | Sierra Nevada (Granada) | 03º 23´ 05" W | 37º 03´ 51" | 2896 | WFC |
| 12 | La Hita (Toledo) | 03º 10' 59" W | 39º 34º 05" | 674 | WFC |

Table 2. Magnitude distribution of Draconids on Oct. 8-9, 2011.

| Method | Number | -4 | -3 | -2 | 0 | +1 | +2 | +3 | +4 | +5 | r |
|---|---|---|---|---|---|---|---|---|---|---|---|
| Visual | 396 | 4 | 5 | 12 | 34 | 70 | 97 | 123 | 48 | 3 | 2.3±0.2 |
| Video | 75 | 1 | 2 | 3 | 6 | 13 | 20 | 27 | 3 | - | |

Table 3. ZHR and flux estimations.

| Interval (UT) | $\lambda_o$ (º) | Number of meteors | ZHR | $\varepsilon$ | Flux ($\times 10^{-3}$ km$^{-2}$ h$^{-1}$) | $\varepsilon_{flux}$ | r | $\varepsilon_r$ |
|---|---|---|---|---|---|---|---|---|
| 19h00-19h15 | 194.99013 | 12 | 127 | 37 | 34 | 10 | 3.2 | 0.6 |
| 19h15-19h30 | 195.00038 | 33 | 223 | 39 | 60 | 10 | 2.1 | 0.5 |
| 19h30-19h45 | 195.01063 | 51 | 419 | 59 | 113 | 16 | 2.3 | 0.7 |
| 19h45-20h00 | 195.02088 | 42 | 259 | 40 | 70 | 11 | 2.0 | 0.7 |
| 20h00-20h15 | 195.03113 | 63 | 371 | 47 | 100 | 13 | 1.8 | 0.6 |
| 20h15-20h30 | 195.04138 | 39 | 235 | 38 | 64 | 10 | 1.9 | 1.2 |
| 20h30-20h45 | 195.05163 | 21 | 230 | 54 | 62 | 15 | 3.5 | 1.2 |
| 20h45-21h00 | 195.06188 | 30 | 269 | 49 | 73 | 13 | 2.7 | 0.8 |
| 21h00-21h15 | 195.07213 | 45 | 394 | 59 | 106 | 16 | 2.1 | 0.7 |
| 21h15-21h30 | 195.08238 | 24 | 220 | 45 | 59 | 12 | 2.0 | 1.2 |
| 21h30-22h00 | 195.09775 | 12 | 50 | 14 | 14 | 4 | 2.0 | 1.4 |
| 22h00-22h30 | 195.11825 | 21 | 129 | 28 | 35 | 8 | 2.4 | 1.7 |

Table 4. Backscatter radio counts obtained by Diego Rodríguez in the observing interval discussed here. In bold are the high rates that reveal dust trail crossing discussed in the text.

| Interval (UT) | Counts | Interval (UT) | Counts | Interval (UT) | Counts | Interval (UT) | Counts | Interval (UT) | Counts |
|---|---|---|---|---|---|---|---|---|---|
| 14:10-14:20 | 4 | 17:10 | 11 | 20:10 | **39** | 23:10 | 3 | 2:10 | 1 |
| 14:20 | 2 | 17:20 | 6 | 20:20 | 22 | 23:20 | 4 | 2:20 | 2 |
| 14:30 | 2 | 17:30 | 5 | 20:30 | 21 | 23:30 | 6 | 2:30 | 4 |
| 14:40 | 2 | 17:40 | **14** | 20:40 | 11 | 23:40 | 4 | 2:40 | 3 |
| 14:50 | 5 | 17:50 | **6** | 20:50 | **16** | 23:50 | 2 | 2:50 | 2 |
| 15:00 | 1 | 18:00 | **11** | 21:00 | **19** | 0:00 | 4 | 3:00 | 3 |
| 15:10 | 6 | 18:10 | 9 | 21:10 | **15** | 0:10 | 0 | 3:10 | 4 |
| 15:20 | 4 | 18:20 | 4 | 21:20 | 5 | 0:20 | 5 | 3:20 | 0 |
| 15:30 | 2 | 18:30 | 6 | 21:30 | 9 | 0:30 | 7 | 3:30 | 6 |
| 15:40 | 7 | 18:40 | 9 | 21:40 | 15 | 0:40 | 8 | 3:40 | 1 |
| 15:50 | 2 | 18:50 | 5 | 21:50 | 9 | 0:50 | 5 | 3:50 | 8 |
| 16:00 | 2 | 19:00 | 19 | 22:00 | 5 | 1:00 | 1 | 4:00 | 4 |
| 16:10 | 3 | 19:10 | 11 | 22:10 | 12 | 1:10 | 5 | 4:10 | 3 |
| 16:20 | 3 | 19:20 | 13 | 22:20 | 6 | 1:20 | 6 | 4:20 | 5 |
| 16:30 | 6 | 19:30 | **29** | 22:30 | 7 | 1:30 | 5 | 4:30 | 5 |
| 16:40 | 7 | 19:40 | **25** | 22:40 | 6 | 1:40 | 4 | 4:40 | 5 |
| 16:50 | 7 | 19:50 | 20 | 22:50 | 3 | 1:50 | 2 | 4:50 | 6 |
| 17:00 | 3 | 20:00 | **26** | 23:00 | 0 | 2:00 | 5 | 5:00-5:10 | 2 |

Table 5. Trajectory, radiant, and velocity data for the 16 high-precision Draconids reduced so far. Equinox (2000.0)

| SPMN Code | $M_v$ | $H_b$ | $H_{max}$ | $H_e$ | $\alpha_g$ (°) | $\delta_g$ (°) | $V_\infty$ | $V_g$ | $V_h$ |
|---|---|---|---|---|---|---|---|---|---|
| 183440 | -1 | 95.2 | 90.3 | 87.7 | 263.7±0.3 | 55.3±0.3 | 23.6±0.3 | 20.95 | 38.94 |
| 184038 | -2 | 95.7 | 90.5 | 87.1 | 263.48±0.14 | 55.55±0.16 | 23.4 | 20.74 | 38.88 |
| 185050 | -5 | 96.4 | 85.3 | 81.7 | 266.9±0.4 | 58.2±0.4 | 24.5 | 21.94 | 39.27 |
| 185948 | -3 | 102.3 | 92.7 | 85.6 | 263.6±0.3 | 55.9±0.3 | 23.8 | 21.18 | 39.18 |
| 191104 | -4 | 96.7 | 87.6 | 78.6 | 262.7±0.3 | 55.6±0.3 | 23.8 | 21.18 | 39.12 |
| 191929 | -2 | 97.9 | 92.3 | 88.5 | 262.65±0.14 | 55.57±0.14 | 23.5 | 20.88 | 39.11 |
| 192250 | -6 | 98.5 | 93.7 | 89.5 | 258.9±0.4 | 54.2±0.4 | 22.9 | 20.19 | 38.67 |
| 192840 | -1 | 94.3 | 84.2 | 83.6 | 263.6±0.3 | 54.5±0.3 | 23.2 | 20.5 | 39.1 |
| 194759 | -11 | 107.3 | 99.1 | 77.1 | 264.15±0.14 | 54.69±0.14 | 23.3 | 20.68 | 39.02 |
| 195157 | -4 | 95.4 | 92.3 | 88.5 | 258.19±0.11 | 55.15±0.14 | 23.2 | 20.57 | 38.71 |
| 201354 | -4 | 93.9 | 89.5 | 85.2 | 268.9±0.4 | 55.81±0.06 | 23.0 | 20.32 | 38.93 |
| 201440 | -3 | 103.7 | 98.4 | 93.1 | 261.14±0.14 | 55.87±0.08 | 23.2 | 20.55 | 38.86 |
| 201453 | -2 | 91.4 | 89.1 | 87.4 | 268.90±0.18 | 56.49±0.05 | 23.0 | 20.29 | 38.73 |
| 201849 | -4 | 92.1 | 88.9 | 82.6 | 259.6±0.5 | 53.58±0.14 | 22.9 | 20.22 | 38.83 |
| 203103 | -3 | 92.8 | 87.5 | 86.8 | 263.2±0.4 | 55.30±0.03 | 23.6 | 21.03 | 38.18 |
| 204801 | -4 | 104.1 | 85.9 | 81.4 | 257.6±0.3 | 54.9±0.3 | 23.5 | 20.91 | 38.87 |
| Average | - | 97.7 | 90.5 | 85.1 | 263.0±0.4 | 55.3±0.3 | 23.4 | 20.76 | 38.90 |

Table 6. Predicted radiant positions and averaged geocentric velocity ($V_g$) of members of 21P dust trails according to Maslov (2011), Vaubaillon et al. (2011), Jenniskens and Vaubaillon (2011). Equinox (2000.00).

| Trail/source | RA (º) | DEC (º) | $V_g$ (km/s) | Source |
|---|---|---|---|---|
| 1900 | 263.3 | +55.8 | 20.9 | Maslov (2011) |
| 1873-1894 | 263.3 | +55.4 | - | Jenniskens and Vaubaillon (2011) |
| SPMN | 263.0±0.4 | 55.3±0.3 | 20.76 | This work (Table 5) |
| AGD campaign | 262.8±0.7 | +55.5±1.1 | 20.98±0.95 | Langbroek (2011) |

Table 7. Orbital elements of the **16** Giacobinid meteors. Equinox (2000.00).

| SPMN Code | Day | q (AU) | a (AU) | e | i (º) | ω (º) | Ω (º) |
|---|---|---|---|---|---|---|---|
| 183440 | 8.77407986 | 0.99688±0.00018 | 3.42±0.25 | 0.709±0.021 | 31.9±0.4 | 185.98±0.24 | 194.97317 |
| 184038 | 8.77821759 | 0.99601±0.00008 | 3.53±0.17 | 0.718±0.013 | 31.3±0.3 | 172.98±0.11 | 194.97727 |
| 185050 | 8.78530324 | 0.99843±0.00013 | 3.80±0.25 | 0.737±0.017 | 33.5±0.3 | 176.8±0.3 | 194.98422 |
| 185948 | 8.79152894 | 0.99671±0.00019 | 3.7±0.3 | 0.729±0.022 | 32.1±0.4 | 186.14±0.24 | 194.99040 |
| 191104 | 8.79934722 | 0.99684±0.00018 | 3.6±0.3 | 0.724±0.022 | 32.2±0.4 | 185.99±0.11 | 194.99811 |
| 191929 | 8.80520486 | 0.99550±0.00010 | 3.6±0.3 | 0.72±0.03 | 31.5±0.4 | 172.46±0.14 | 195.0039 |
| 192250 | 8.80752546 | 0.9928±0.0004 | 3.2±0.3 | 0.69±0.03 | 30.6±0.5 | 169.9±0.4 | 195.00623 |
| 192840 | 8.81157407 | 0.9964±0.0002 | 3.6±0.2 | 0.72±0.02 | 30.8±0.3 | 173.4±0.2 | 195.0102 |
| 194759 | 8.82498843 | 0.99672±0.00007 | 3.7±0.3 | 0.731±0.021 | 31.1±0.4 | 186.12±0.13 | 195.02346 |
| 195157 | 8.82774769 | 0.99190±0.00017 | 3.20±0.22 | 0.692±0.020 | 31.2±0.6 | 169.21±0.19 | 195.02610 |
| 201354 | 8.84299421 | 0.9987±0.00015 | 3.41±0.23 | 0.71±0.03 | 30.8±0.3 | 177.5±0.3 | 195.04125 |
| 201440 | 8.84351968 | 0.99519±0.00011 | 3.21±0.19 | 0.684±0.020 | 31.2±0.3 | 172.02±0.21 | 195.04174 |
| 201453 | 8.84367361 | 0.9988±0.0003 | 3.22±0.21 | 0.690±0.020 | 31.0±0.6 | 177.71±0.11 | 195.0418 |
| 201849 | 8.84640046 | 0.9931±0.0003 | 3.3±0.3 | 0.700±0.021 | 30.5±0.3 | 170.2±0.4 | 195.0446 |
| 203103 | 8.85490046 | 0.9900±0.0005 | 2.79±0.15 | 0.645±0.019 | 32.8±0.4 | 167.6±0.4 | 195.0529 |
| 204801 | 8.86667245 | 0.9917±0.0003 | 3.34±0.24 | 0.704±0.021 | 31.7±0.4 | 169.2±0.3 | 195.06459 |
| Average | - | 0.9954±0.0003 | 3.40±0.23 | 0.705±0.021 | 31.5±0.4 | 174.86±0.23 | - |

Table 8. Disruption heights. velocity at disruption point. Atmospheric US standard density from which the dynamic strengths of selected Draconids are computed.

| SPMN code | Magnitude | $H_{max}$ (km) | v (km/s) | $\rho$ ($\times 10^{-9}$ g/cm$^3$) | $\tau$ ($\times 10^2$ dyn/cm$^2$) |
|---|---|---|---|---|---|
| 185050 | -5 | 85.3 | 20.1 | 6.3528 | 35±1 |
| 191104 | -4 | 87.6 | 20.4 | 4.1700 | 25±1 |
| 194759 | -11 | 99.1 | 22.5 | 0.3855 | 1.9±0.1 |
| 204801 | -4 | 85.9 | 21.2 | 5.7007 | 35±1 |

# **FIGURES**

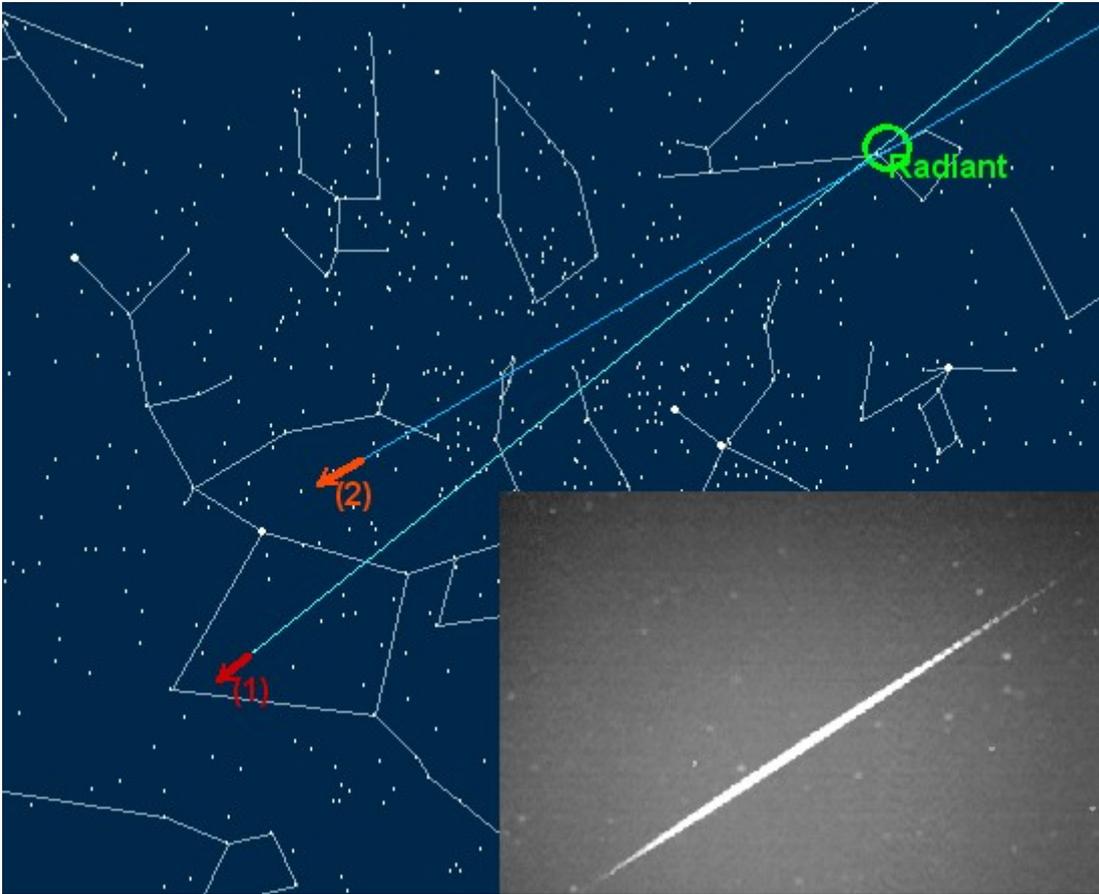

Figure 1. Accurate astrometric measurements of Giacobinid meteors in reference with background stars allow us to infer their respective radiants. On the bottom-left border a -2 meteor recorded at 20h54m52s UTC from Seville [2] SPMN station. The stellar chart shows the event from two stations and its apparent radiant derivation.

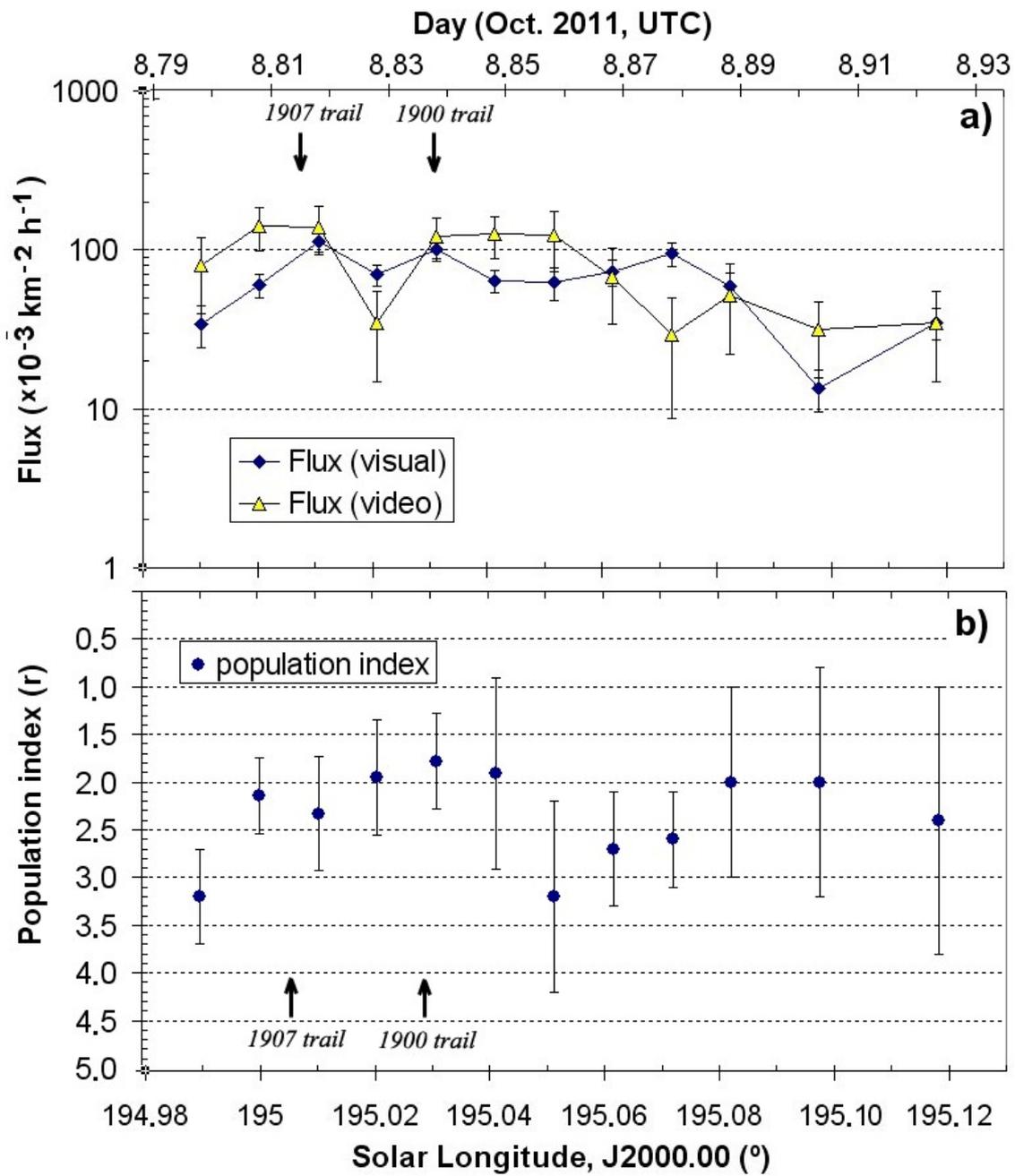

Figure 2. a) Derived meteoroid fluxes, and b) Population index values for visual and video data.

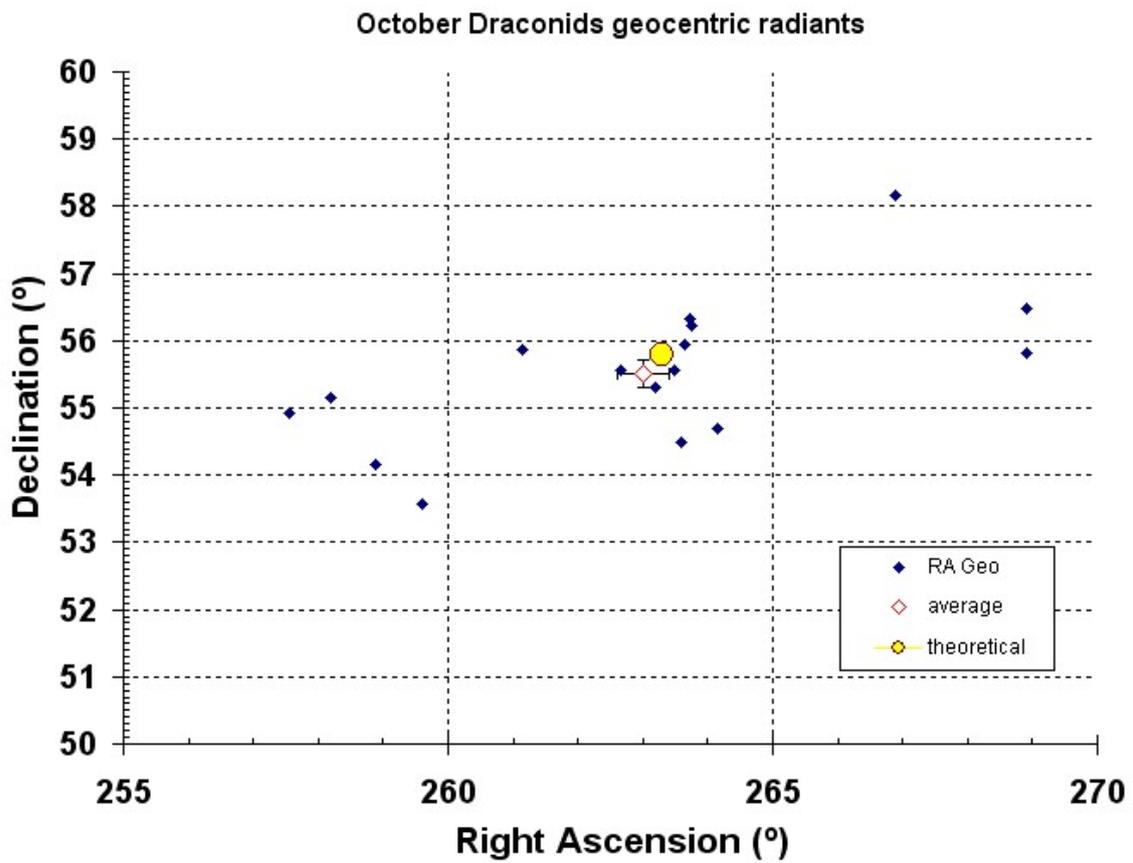

Figure 3. Computed Giacobinid geocentric radiants, and our averaged radiant position compared with the theoretical position given by Maslov (2011).

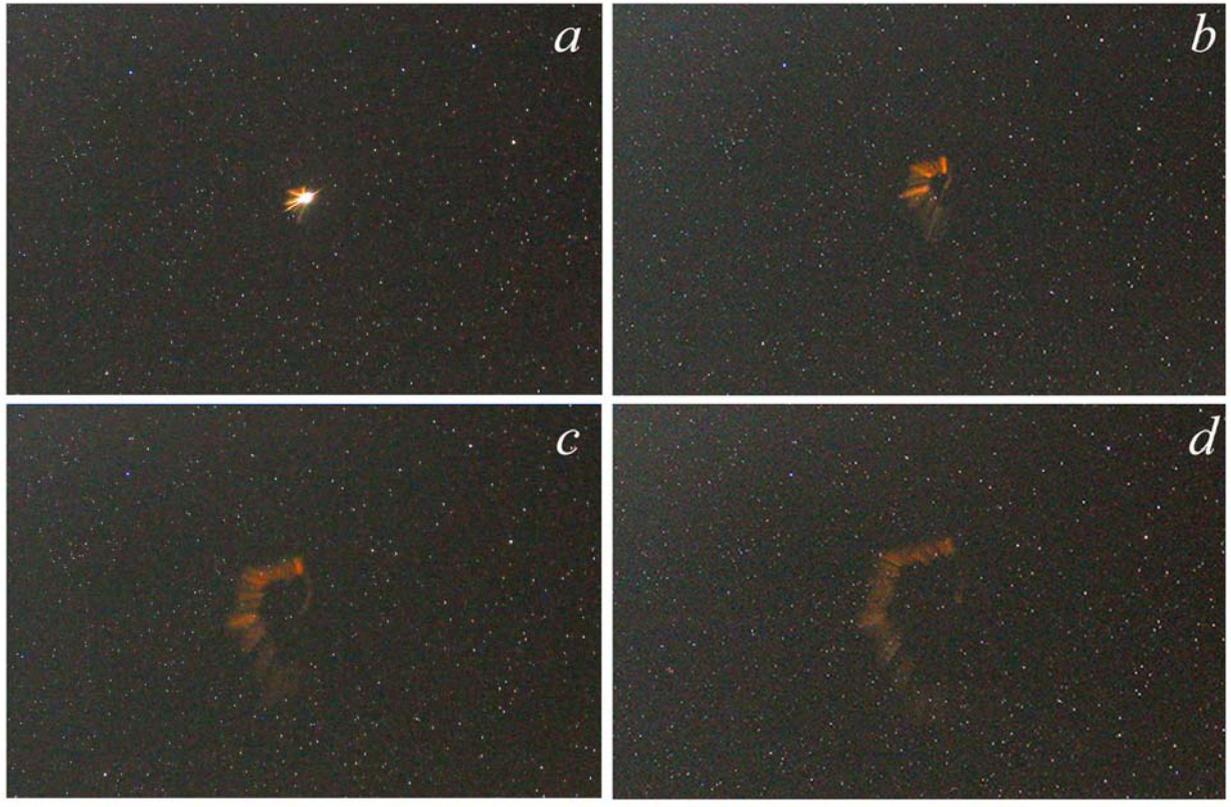

**Figure 4**. Image sequence taken by Antonio Francisco Marín of an almost stationary –10.5 magnitude SPMN194759 bolide seen from El Picacho (Cádiz), and its persistent train left behind. Propagation is nicely seen in the 30 seconds exposure consecutive images. First picture taken at 19h47m50s UTC, and readout time between images of about 3 seconds.

21